# Author-level metrics in the new academic profile platforms: The online behaviour of the Bibliometrics community

Alberto Martín-Martín[1], Enrique Orduna-Malea[2], Emilio Delgado López-Cózar[1]*

[1] Facultad de Comunicación y Documentación, Universidad de Granada, Colegio Máximo de Cartuja s/n, 18071, Granada, Spain.
[2] Universitat Politècnica de València, Camino de Vera s/n, 46022, Valencia, Spain.
* Corresponding author: E-mail address: edelgado@ugr.es (E. Delgado).

Abstract

The new web-based academic communication platforms do not only enable researchers to better advertise their academic outputs, making them more visible than ever before, but they also provide a wide supply of metrics to help authors better understand the impact their work is making. This study has three objectives: a) to analyse the uptake of some of the most popular platforms (Google Scholar Citations, ResearcherID, ResearchGate, Mendeley and Twitter) by a specific scientific community (bibliometrics, scientometrics, informetrics, webometrics, and altmetrics); b) to compare the metrics available from each platform; and c) to determine the meaning of all these new metrics. To do this, the data available in these platforms about a sample of 811 authors (researchers in bibliometrics for whom a public profile Google Scholar Citations was found) were extracted. A total of 31 metrics were analysed. The results show that a high number of the analysed researchers only had a profile in Google Scholar Citations (159), or only in Google Scholar Citations and ResearchGate (142). Lastly, we find two kinds of metrics of online impact. First, metrics related to connectivity (followers), and second, all metrics associated to academic impact. This second group can further be divided into usage metrics (reads, views), and citation metrics. The results suggest that Google Scholar Citations is the source that provides more comprehensive citation-related data, whereas Twitter stands out in connectivity-related metrics.

## 1. Introduction

Last decade has witnessed the emergence of a plethora of new communication channels and social collaboration platforms where academic outputs are susceptible of being indexed, searched, located, read, and mentioned (Priem & Hemminger, 2010; Piwowar, 2013). The degree to which these online platforms are used (as well as the metrics they offer) provide new insights about the current dynamics of research activity, not without introducing some methodological concerns (Bornmann, 2014; 2016; Sugimoto et al., 2017).

## 1.1. Online academic communication channels: from personal websites to academic profiles

Academic personal websites are probably the first venues where scholars started to disseminate their personal information, current activities and projects, and their lists of academic contributions. Scientists' personal websites have been extensively studied from a webometric approach (Barjak, Li & Thelwall, 2007; Mas-Bleda & Aguillo, 2013; Más-Bleda et al, 2014). The literature on this issue describes significant differences in web visibility according to disciplines, countries, gender, and age. Additionally, several studies find an overall low presence (number of researchers with personal website) and lack of essential information in these websites (Chen et al, 2009; Mas-Bleda & Aguillo, 2013).

However, despite their initial important role as the first venues where researchers could make their work available to others, personal websites did not allow much interaction between researchers, nor were they suitable tools to record and provide metrics about the authors' academic contributions. Online academic profile platforms are now filling this gap. These online environments usually supply a variety of metrics that capture the diversity of actions and interactions that can occur amongst scientists in the digital space (Haustein, 2016), actions that already contribute to reshape the scholarly reputation of authors (Jamali, Nicholas & Herman, 2015).

Naturally, the features available in these platforms vary slightly from one to another (Jordan, 2014a), but they usually provide researchers with the choice to create an academic profile, and upload their research outputs (not only published materials, but also posters, presentations, software, and other kinds of unpublished materials). These outputs can then be accessed by other researchers, who can download them or comment them. Researchers can interact in other ways with these platforms (tagging and following profiles, asking and answering questions). The most popular academic platforms that provide social features and author profiles are ResearchGate (Kadriu, 2013; Thelwall & Kousha, 2015; Nicholas, Clark & Herman, 2016), Academia.edu (Thelwall & Kousha, 2014), and Mendeley (Li, Thelwall & Giustini, 2011; Mohammadi & Thelwall, 2014). Additionally, academic databases have also developed platforms that enable authors to create profiles to list their publications, based on the coverage of each platform. Among these, we can find ResearcherID (Web of Science), Scopus ID (Scopus), and Google Scholar Citations (GSC) (Google Scholar).

The increasing use of all these social platforms by scholars was reported in the results of a survey carried out by the Nature Publishing Group (Van Noorden, 2014). The data collected in this survey was also made openly available (Nature Research, 2014). Jordan (2014b) re-used this dataset and found significant differences in the perceived usefulness of social network platforms depending on the respondents. Of the 480 researchers in the Humanities, Arts, and Social Sciences who responded, more than 70% declared that they were aware of Google Scholar (either the search engine or the profile service) and visited it regularly. This figure decreased to 61% for the respondents from Science and Engineering (n ~ 3,000).

Mas-Bleda et al. (2014) analysed the presence of 1,517 highly-cited European researchers in several platforms (GSC, Microsoft Academic Search, Mendeley, Academia.edu, Linkedin, and SlideShare). She found that the use of online academic profile services by these top-cited researchers was still low (only 9% of the researchers in the sample had a public profile in GSC). Additionally, this study also reported high inter-disciplinary differences (24% of the social scientists had a profile created in GSC; 8% in Mendeley). This study did not consider ResearchGate due to the low number of researchers in that platform at the time. However, ResearchGate later proved to be the most used academic profile platform according to a survey about the use of academic communication tools carried out by Kramer and Bosman (2016). According to this survey, which was responded by over 20,000 people related to academia (mostly researchers, but also librarians, publishers, and people from the industry and government), the most used academic profile platform overall was ResearchGate (66%), followed by GSC (62%).

**1.2. Online academic metrics: from citations to author-level metrics**

Since the classic study by Bollen et al. (2009), where the data came primarily from usage logs provided by publishers, many papers have been published on the nature of online article-level metrics. Some of these works intended to shed light on the correlation between traditional citation-based metrics and the flourishing array of altmetrics across disciplines and platforms (Priem, Piwowar & Hemminger, 2012; Thelwall et al, 2013; Costas, Zahedi & Wouters, 2015; De Winter, 2015). Nevertheless, prudence has been advised when interpreting the meaning of these correlations (Thelwall, 2016).

Similar studies, but focused on author-level metrics, instead of article-level metrics, are not as frequent (Bar-Ilan et al, 2012; Wildgaard, Schneider & Larsen, 2014; Orduna-Malea, Martín-Martín & Delgado López-Cózar, 2016a). The few studies on this front have found that the author-level metrics in a given platform tend to correlate with one another. This may be related to the claim that authors primarily create these profiles to advertise themselves and not to collaborate, in line with the Diogenes Club analogy proposed by Ortega (2016a). On the other hand, reports of high correlations between metrics across different platforms have been scarce to date.

Some earlier studies have addressed the similarity of the metrics reported by GSC to those displayed by other platforms. Ortega (2015b) analysed researchers working at the Spanish National Research Council (CSIC), and Mikki et al. (2015) chose to study researchers at the University of Bergen. Despite using very different samples, the results of these two studies agree on the low correlations found between citation-based indicators and altmetrics (based on social interactions) at the level of authors. Nevertheless, these studies were limited to authors from two particular institutions.

**1.3. Towards a disciplinary study of author-level-metrics**

At the discipline level, Bar-Ilan et al. (2012) sampled 57 presenters at the 2010 Leiden STI Conference (an event that is mainly devoted to discuss issues in the field of bibliometrics and related areas), collecting publication and citation counts from a variety of web platforms. The authors found that 70% of presenters had a LinkedIn account, 23% of them had public GSC profiles, and 16% of them were on Twitter, as of 2012. Later, Haustein et al. (2014) studied the same sample of authors, finding that 58% (33) of the authors had a profile in ResearchGate, and 53% (30) had a public profile in GSC.

However, an exhaustive multi-platform study that analyses an entire academic discipline has not been carried out yet. It is reasonable to assume that, given that each platform has its own specific userbase and document coverage (which inevitably affects the metrics provided by the platform), none of these platforms, by themselves, are able to provide a complete and accurate portrayal of an author's impact. For this reason, we believe a multi-platform approach is necessary if the goal is to collect all the available evidence of an author's impact.

By working with an homogeneous sample (researchers that work on the same discipline), we believe we might find existing patterns regarding the use of public academic profiles, and shed light on the question of the meaning of the metrics that are displayed in these profiles.

**1.4. Research questions**

The main goals of this study are to analyse the uptake of some of the most popular platforms (Google Scholar Citations, ResearcherID, ResearchGate, Mendeley and Twitter) by a specific scientific community (bibliometrics, scientometrics, informetrics, webometrics, and altmetrics), to compare the metrics available from each platform, and to determine the meaning of all these new metrics. The reason behind the selection of this discipline is that this discipline is the one the authors know best, and expertise in the field is in this case necessary to find patterns that arise from the metrics that will be analysed.

The following research questions (RQ) are proposed:

> (RQ1): Is there a large enough volume of data at the author-level (author-level metrics) to allow the development of bibliometric analysis? Are there significant differences in the quantity of data available across platforms?
> (RQ2): Which are the main dimensions of online impact captured by author-level metrics? Is there any significant correlation between the different author-level metrics offered by the social platforms analysed when considering authors that belong to one specific academic discipline?

# 2. Methods

The construction of the data sample followed the MADAP (Multifaceted Analysis of Disciplines through Academic Profiles) method (Martín-Martín, Orduna-Malea & Delgado López-Cózar, 2018), which consisted basically of four steps: identification of authors, location of academic profiles for those authors, extraction of author-level metrics from each platform, and, lastly, statistical data analysis.

**2.1. Identification of authors**

We first established two required criteria to include authors in the sample:

a) Authors who have published in the areas of bibliometrics, scientometrics, informetrics, webometrics or altmetrics, and
b) Authors who have created a GSC public profile.

Regarding the first criterion, authors from all these subfields were included, because all of them are related, to a greater or lesser extent, with the quantitative studies of science.

The second criterion was established mainly because a previous study (Haustein et al., 2014) had established that GSC was the most popular platform among researchers in this discipline. Bosman and Kramer (2016) confirmed that GSC and ResearchGate were the most widely used academic profile platforms.

In order to identify the corpus of authors, three complementary procedures were carried out, described below.

All queries and data extraction were performed on July 24th 2015.

*Method 1: topical keywords*

We obtained a list of frequently-used descriptive terms in the discipline. To do this, the bibliographic records from all indexed articles published in an initial seed of five core journals exclusively devoted to the discipline were automatically retrieved using the Web of Science and Scopus. This set of sources was composed by the Journal of Informetrics, Research Evaluation and Scientometrics (the three journals with a declared scope mainly devoted to the field with higher impact factor in the Journal Citation Reports, 2016 edition). In addition to this, Cybermetrics (in order to cover the Webometrics subfield) and ISSI Conference Proceedings (in order to cover the major worldwide conference on the topic) were also included as initial sources seeds.

Next, all significant terms from the documents' titles and keywords (when available) were extracted and analysed to find out their frequency of use. This vocabulary was later cleaned, merging variants (for example, scientometric and scientometrics), deleting duplicates, and excluding generic terms that did not unequivocally describe the discipline or its main topics (e.g., credit, item, program, content, etc.). All the profiles in GSC that included at least one of these descriptive keywords (see Table 1) was included in the sample. Keyword variants (such as misspelled words, the same terms in other languages, etc.) were also considered. Terms not included as GSC keywords in any profile were excluded.

*Method 2: Additional searches in Google Scholar*

In order to capture authors relevant to the discipline but who have not included any of the relevant terms selected in method 1 or do not belong to any of the institutions mentioned in method 2, we also conducted a series of searches in Google Scholar (the search engine) with the intention to find these stragglers. Two types of searches were carried out: a) keyword searches using the terms obtained in method 1 (18 queries in total); journal name searches, to find obtain the documents published in the journals mentioned in method 1 (5 queries in total). The keyword searches were limited so that only documents with those terms in the titles were retrieved (to increase precision).

All the results to these queries (a maximum of 1,000 results per query, as per Google Scholar's limitations) were extracted. Then, although the information provided by GS is not always complete, we extracted all the instances of profile URLs that are displayed in the list of authors of each article (see Figure 1). All the profiles that had already been identified in previous steps were removed. Given how Google Scholar ranks results, which is basically according to the number of citations a document has received (Martín-Martin et al, 2017), we are reasonably confident that extracting the first 1,000 results of each query is enough to identify most of the relevant authors we might have been missing in the previous steps.

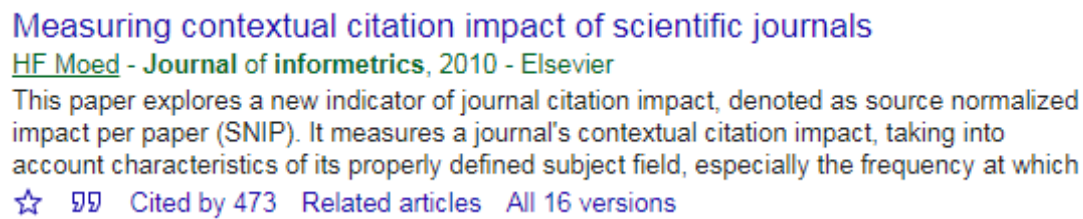

**Fig. 1.** Example of a Google Scholar bibliographic record.

**Table 1.**
List of terms used as author keywords in GSC profiles representing the field of bibliometrics.

| Topic Keywords | |
|---|---|
| Altmetrics | Research Assessment |
| Bibliometrics | Research Evaluation |
| Citation Analysis | Research Policy |
| Citation Count | Science and Technology Policy |
| H-Index | Science Evaluation |
| Impact Factor | Science Policy |
| Informetrics | Science Studies |
| Patent Citation | Scientometrics |
| Quantitative Studies of Science and Technology | Webometrics |

*Method 3: Institutional affiliation*

As a supplementary search strategy, we considered searching authors affiliated to centres or departments that produce research in the area of bibliometrics, regardless of the research keywords used by authors in their public profiles. To do this, we obtained the list of researchers from the websites of these centers and searched them manually. The research centres we considered were: CWTS (Centre for Science and Technology Studies) (cwts.leidenuniv.nl; cwts.nl), Cybermetrics Lab (webometrics.info; ipp.csic.es), DZHW (Deutsches Zentrum für Wissenschafts- und Hochschulforschung) (dzhw.eu), ECOOM (Expertisecentrum Onderzoek en Ontwikkelingsmonitoring) (ecoom.be), EC3 Research Group: Evaluación de la Ciencia y de la Comunicación Científica (ec3.ugr.es), Science-Metrix (science-metrix.com), Scimago (scimagojr.com; scimagolab.com), SciTech Strategies (scitechstrategies.com), Statistical Cybermetrics Research Group (cybermetrics.wlv.ac.uk). However, this method did not add any additional authors to our sample, and therefore, the list of centers were not exhaustively expanded and this method was not ultimately used. However, it may be useful as a complement search in other fields.

Since GSC gives authors complete control over how their profile is set (personal information, institutional affiliation, research interests, as well as their scientific production), a systematic manual revision was carried out in order to:

- Detect false positives: authors whose scientific production doesn't have anything to do with this discipline, even though they labelled themselves with one or more of the keywords associated with it.

- Classify authors in two categories:

a) *Specialists*: authors whose scientific production substantially falls within the field of bibliometrics.
b) *Occasional*: authors who have sporadically published bibliometric studies, or whose field of expertise is closely related to Scientometrics (social, political, and economic studies about science), and therefore they can't be strictly considered bibliometricians.

We decided to consider as specialists those who meet the following criterion: at least half of the documents which contribute to their h-index in their GSC profile should fall within the limits of the field of bibliometrics.

To help delimiting the limits of the field, we considered not only the titles of the documents but also the venue where they had been published. To do this, we first defined a Bradford-like core of journals about bibliometrics (Scientometrics, Journal of Informetrics, JASIST, Research Evaluation, Research Policy, and Cybermetrics), followed by other LIS journals which also publish numerous bibliometric studies (Journal of Information Science, Information Processing & Management, Journal of Documentation, College Research & Libraries, Library Trends, Online Information Review, Revista Española de Documentación Científica, Aslib Proceedings, and El Profesional de la Información). Lastly, journals devoted to social and political studies about science (Social Studies of Science, Science and Public Policy, Minerva, Journal of Health Services Research Policy, Technological Forecasting and Social Change, Science Technology Human Values, Environmental Science Policy, and Current Science) were also searched. The selection of these journals was mainly based on our expert judgement, which also matches to a large degree the empirical results obtained by Hood and Wilson (2001), who analysed the bibliometrics literature and reported a list of the most productive journals in the field of bibliometrics.

In the end, we gathered a total of 811 GSC profiles, out of which the 48.8% (396) were classified as specialists, and the remaining 51.2% (415) as occasional authors.

**2.2. Searching in other platforms**

In addition to GSC, the academic profile services we considered in this analysis were ResearcherID[1], ResearchGate[2], and Mendeley[3]. Other academic profile services were not considered due to several reasons. Preliminary explorations showed that there was a very low coverage of authors in the discipline in platforms like Academia.edu and Loop. AMiner was discarded because it was found to be outdated, and Microsoft Academic was still in beta when this study was carried out.

We decided to include Twitter[4] because, although this platform is not designed to set up academic profiles, participation in this platform affects the level of dissemination of research papers (Ortega, 2016b), thus capturing an important dimension of an author's online visibility.

---

[1] https://www.researcherid.com
[2] https://www.researchgate.net
[3] https://www.mendeley.com
[4] https://twitter.com

## 2.3. Obtaining the metrics

For each of the 811 authors in our sample, we manually checked whether they had also created profiles in ResearcherID, ResearchGate, Mendeley, and Twitter, by searching their names in each of the profile services' search features, and by searching their names in Google in combination with the name of the profile platform. In the case of Twitter, additional searches through well identified authors' followers were used in order to find authors whose profile names did not correspond with their personal names.

When a profile was found, all available author-level metrics were extracted. Custom automated parsers were developed for this purpose. The data collection in these platforms was carried out between the 4$^{th}$ and the 10$^{th}$ of September, 2015.

A total of 31 author-level metrics were extracted from GSC and the rest of profile platforms. These were the metrics that were available in each platform at the time of data collection. Some platforms might now offer new metrics, or they might have stopped displaying some of the metrics we discuss in this analysis. Their scope and definition can be found in Table 2. Additionally, we categorize each metric according to its nature: total (size-dependent; 22 metrics), average (size-independent; 3 metrics), and hybrid (composite indicator; 6 metrics).

**Table 2.**
List of Author-Level metrics.

| PLATFORM | METRIC | DEFINITION | CATEGORY |
|---|---|---|---|
| **GSC** | Citations | Number of citations to all publications. Computed for citations from all years, and citations received since 2010 | Total |
| | h-index | The largest number h such that h publications have at least h citations. Computed for citations from all years, and citations received since 2010 | Hybrid |
| | i10 index | Number of publications with at least 10 citations. Computed for citations from all years, and citations received since 2010 | Total |
| **RESEARCHER ID** | Total Articles | Number of items in the publication list | Total |
| | Articles with Citation Data | Only articles added from *Web of Science Core Collection* can be used to generate citation metrics, even though the publication list may contain articles from other sources. This value indicates how many articles from the publication list were used to generate the metrics | Total |
| | Sum of Times Cited | Total number of citations to any of the items in the publication list from *Web of Science Core Collection*. The number of citing articles may be smaller than the sum of the times cited because an article may cite more than one item in the set of search results | Total |
| | Average Citations per Item | Average number of citing articles for all items in the publication list from *Web of Science Core Collection*. It is the sum of the times cited divided by the number of articles used to generate the metrics | Average |
| | h-index | An author has a h-index of "h" when "h" of its articles has achieved at least "h" citations. | Hybrid |
| **RESEARCH GATE** | RG Score | It is a composite indicator that according to RG measures scientific reputation based on how an author's research is received by his/her peers. The exact method to calculate this metric has not been made public, but it takes into account how many times the contributions (papers, data, etc.) an author uploads to *ResearchGate* are visited and downloaded, and also by whom (reputation) | Hybrid |
| | Publications | Total number of publications an author has added to his/her profile in *ResearchGate* (full-text or no) | Total |
| | Views | Total number of times an author's contributions to *ResearchGate* have been visualized. This was later combined with the "Downloads" metric to form the new "Reads" indicator, but the data collection for this product was made before this change came into effect | Total |
| | Downloads | Total number of times an author's contributions to *ResearchGate* have been downloaded. This metric was later combined with the "Views" indicator to form the new "Reads" indicator, but the data collection for this product was made before this change came into effect | Total |
| | Citations | Total number of citations to the documents uploaded to the profile. | Total |
| | Impact Points | Sum of the JCR impact factors of the journals where the author has published articles. This metric is no longer available in public RG profiles. | Hybrid |
| | Profile views | Number of times the author's profile has been visited. This indicator is no longer public. Currently, users can only see their own profile views count, but not other users'. | Total |
| | Following | Number of *ResearchGate* users the author follows (friends) | Total |
| | Followers | Number of *ResearchGate* users who follow the author | Total |
| **MENDELEY** | Readers | This number represents the total number of times a *Mendeley* user has added a document by this author to his/her personal library | Total |
| | Publications | Number of publications the author has uploaded to *Mendeley* and classified as "My Publications" | Total |
| | Readers per document | Number of Readers divided by the number of publications per each author | Average |
| | Followers | Number of *Mendeley* users who follow the author in *Mendeley* | Total |
| | Following | Number of *Mendeley* users the author follows in *Mendeley* | Total |
| **TWITTER** | Tweets | Total number of tweets an author has published according to his/her profile | Total |
| | Followers | Number of *Twitter* users who follow the tweets published by the author | Total |
| | Following | Number of *Twitter* users the author follows | Total |

| | Days registered | Number of days since the author created his/her account on *Twitter* | Total |
|---|---|---|---|
| | Sum Retweets | Number of Retweets received for the author. These data was extracted using the software Webometric Analyst (Statistical Cybermetrics Research Group, 2011), and is limited to the data that the Twitter API allowed us to extract, meaning that it was not possible to extract all the tweets from all authors, especially those that are more active on Twitter. Therefore, the sum of retweets for these authors are incomplete as well. | Total |
| | H-Retweets | An author has a h-Retweet of "n" when "n" of its tweets has achieved at least "n" Retweets. | Hybrid |

\* This metric is currently available only

### 2.4. Statistical data analysis

Because the goal of this analysis is to find any potential relationships between metrics, this analysis has no preconceptions regarding the nature of each of these metrics. For this reason, the Spearman correlation ($\alpha < 0.05$) was computed for all pairs of the 31 metrics under study, and a Principal Component Analysis (PCA) was applied in order to display by means of two-dimensional axis the relatedness between the variables analysed with the aim to synthetize components whose relatedness may illustrate different web impact dimensions (RQ2). Both for the correlations and the PCA, all the observations with null values were removed.

Correlations are considered useful in high-level exploratory analyses to check whether different indicators reflect the same underlying causes (Sud & Thelwall, 2014). Spearman correlations were used because it is well-known that citation counts and other impact-related metrics are highly skewed (de Solla Price, 1965). The PCA, on the other hand, has been proved as a valid technique to reduce the dimensionality of the dataset through the identification of principal components (Jollife, 1986). In this case, due to the nature of the web data distribution, Spearman similarity (with varimax rotation of axes and uniform weighting to simplify the data interpretation) was applied.

# 3. Results

### 3.1. Online presence of the bibliometrics community

The distribution of authors according to the number of platforms in which they have created a personal profile shows a high degree of social presence (Table 3). Authors with two (26.3%) and three (23.3%) profiles are the more numerous groups whereas there is a small group (11.3%) of authors (most of them specialists) with presence in all five platforms analysed.

**Table 3.**
Social presence (number of authors) of the bibliometrics community.

| NUMBER OF PLATFORMS | AUTHORS | | | |
|---|---|---|---|---|
| | SPECIALIST | OCCASIONAL | TOTAL | % |
| 5 | 58 | 34 | 92 | 11.3 |
| 4 | 82 | 80 | 162 | 20.0 |
| 3 | 83 | 106 | 189 | 23.3 |
| 2 | 99 | 114 | 213 | 26.3 |
| 1 | 74 | 81 | 155 | 19.1 |
| **TOTAL** | **396** | **415** | **811** | |

The use of each specific social platform reveals that ResearchGate is, after GSC, the second most used platform by the authors in our sample (67%), followed at some distance by Mendeley (41.2%). However, the number of Mendeley profiles is misleading, since 17.1% of them are basically empty. ResearcherID profiles suffer from the same issue (34.5% of the profiles are empty). Twitter is the least used platform, since only 33.2% of specialist authors (and 26% of occasional authors) have created a Twitter profile. Additionally, most of the authors in our sample that have presence in Twitter, ResearcherID or Mendeley are specialists, while most of the authors in ResearchGate are only occasional authors in bibliometrics (Table 4).

**Table 4.**
Degree of use of social platforms according to the type of author (specialist and occasional).

| WEB PLATFORMS | AUTHORS | | | | | |
|---|---|---|---|---|---|---|
| | SPECIALIST | % | OCCASIONAL | % | TOTAL | % |
| * GSC | 396 | 100 | 415 | 100 | **811** | 100 |
| ResearcherGate | 260 | 65.7 | 283 | 68.2 | **543** | 67.0 |
| Mendeley | 169 | 42.7 | 165 | 39.8 | **334** | 41.2 |
| ResearcherID | 182 | 46.0 | 146 | 35.2 | **328** | 40.4 |
| Twitter | 132 | 33.3 | 108 | 26.0 | **240** | 30.0 |

* All authors in the sample have a profile in GSC.

The combination of profiles used by the authors in our sample (specialists and occasional) is shown in Figure 2.

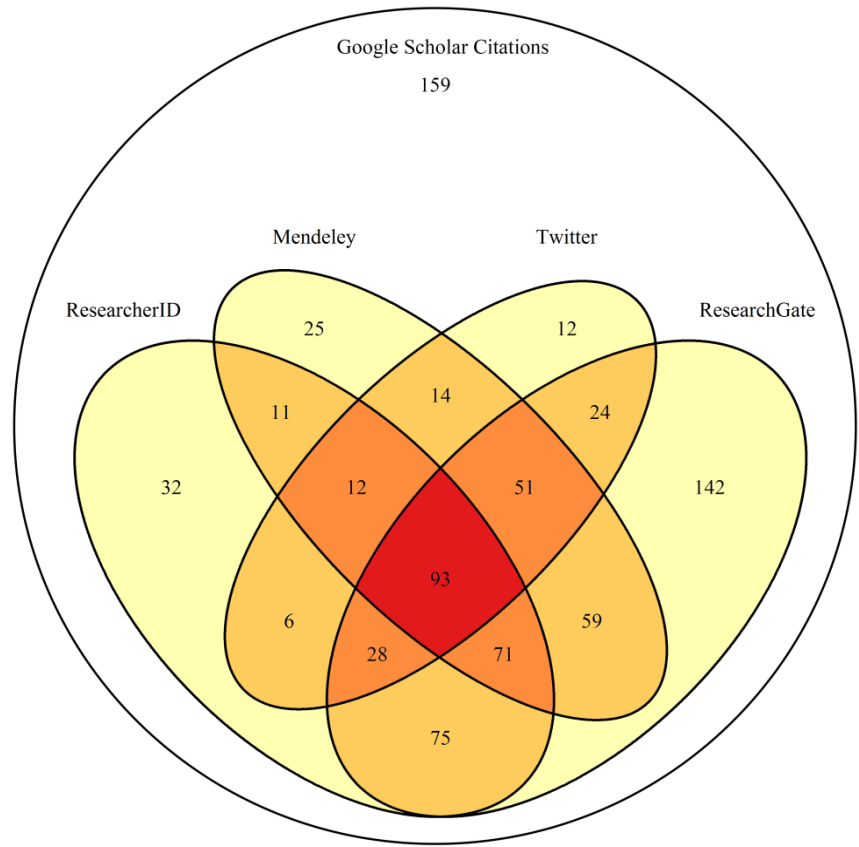

**Fig. 2.** Combination of academic profiles used by the bibliometrics community.

We can observe a great number of researchers who only have a profile in GSC (159) whereas the preferred combination corresponds to GSC and ResearchGate (142). Of the four platforms that we studied (other than Google Scholar Citations), ResearchGate is the one with the highest uptake among the authors in our sample (543 authors had a profile in this platform 66% of the sample). The remaining combinations seem to be more unusual. For example, there are only 12 authors who use only GSC and Twitter or only 11 authors who use only GSC, ResearcherID, and Mendeley.

As it was previously stated, the are many available venues where authors can showcase their work and themselves. As is natural, each author has his/her own preferences, and as a consequence, each profile service offers a different array of products (authors). This issue can be observed just by considering the top 5 authors according to each of the metrics (Table 5). While GSC, ResearcherID, and ResearchGate (all more academic-oriented) seem to portray a similar picture of the discipline, Mendeley, and particularly Twitter, provide a quite different snapshot.

**Table 5.**
Top 5 Author performers according to each of the metrics in each of the academic profiles.

| GSC | | | | | |
|---|---|---|---|---|---|
| **Citations (5 years)** | **H-Index (5 years)** | **I10 index (5 years)** | **Citations (all)** | **H-Index (all)** | **i10 index (all)** |
| L Leydesdorff | L Leydesdorff | L Leydesdorff | L Leydesdorff | L Leydesdorff | L Leydesdorff |
| M Thelwall | M Thelwall | M Thelwall | E Garfield | M Thelwall | E Garfield |
| E Garfield | W Glänzel | W Glänzel | M Thelwall | E Garfield | M Thelwall |
| W Glänzel | L Bornmann | R Rousseau | DJS Price | W Glänzel | R Rousseau |
| R Rousseau | E Garfield | L Bornmann | F Narin | AF.J. van Raan | W Glänzel |

| RESEARCHER ID | | | | |
|---|---|---|---|---|
| **Articles with cit. data** | **Citations / Item** | **Total Articles** | **Citations** | **h-index** |
| AK Sahu | H Small | AK Sahu | AK Sahu | E Garfield |
| E Garfield | L Meho | E Garfield | E Garfield | AK Sahu |
| L Leydesdorff | I Rafols | HD White | L Leydesdorff | L Leydesdorff |
| W Glänzel | M Meyer | L Leydesdorff | W Glänzel | W Glänzel |
| P Jacso | CS Wagner | F Moya | A Schubert | M Thelwall |

| RESEARCHGATE | | | | | | | | |
|---|---|---|---|---|---|---|---|---|
| **RG Score** | **Impact Points** | **Publications** | **Citations** | **Downloads** | **Views** | **Profile Views** | **Followers** | **Following** |
| L Leydesdorff | L Leydesdorff | L Leydesdorff | L Leydesdorff | L Leydesdorff | L Leydesdorff | NA Ebrahim | NA Ebrahim | NA Ebrahim |
| L Bornmann | L Bornmann | R Rousseau | W Glänzel | NA Ebrahim | M Thelwall | C Chen | L Leydesdorff | G. Rathinasabapathy |
| R Rousseau | R Rousseau | M Thelwall | M Thelwall | C Chen | C Chen | Z Chinchilla | M Thelwall | A Keramatfar |
| M Thelwall | A Schubert | S Darmoni | F Narin | M Thelwall | S Darmoni | M Thelwall | Z Chinchilla | IF Aguillo |
| W Glänzel | M Thelwall | C Chen | A Schubert | F Moya | F Moya | L Leydesdorff | IF Aguillo | OB Onyancha |

| MENDELEY | | | |
|---|---|---|---|
| **Readers** | **Publications** | **Followers** | **Following** |
| M Thelwall | RSJ Tol | H Aziz | H Aziz |
| RSJ Tol | P Mayr | J Pacheco | J Pacheco |
| J Vanclay | J Vanclay | C Neylon | E Romero |
| M Pautasso | M Thelwall | l Michán | C Neylon |
| AW Harzing | IF Aguillo | L Adriaanse | l Michán |

| TWITTER | | | |
|---|---|---|---|
| **Tweets** | **Days registered** | **Followers** | **Following** |
| S Fausto | D Hendrix | J Priem | IF Aguillo |
| A Ramos | J Delasalle | IF. Aguillo | A Ramos |
| D Giustini | Á Cabezas | D Giustini | J Pacheco |
| IF Aguillo | S Konkiel | S Konkiel | NA Ebrahim |
| S Konkiel | K Holmberg | Á Cabezas | Y Milanes |

## 3.2. Data available to generate Author-Level Metrics

Table 6 provides an indication of the volume of data available in each platform, by comparing the median values of similar indicators across different platforms.

**Table 6.**
Median of principal online metrics broken down by category.

| TYPE | SOURCE | MEDIAN |
|---|---|---|
| **Citations** | GSC | 156 |
|  | ResearchGate | 85 |
|  | Researcher ID | 63 |
| **Publications** | ResearchGate | 27 |
|  | Researcher ID | 15 |
|  | Mendeley | 9 |
| **H Index** | GSC | 6 |
|  | Researcher ID | 4 |
| **Followers** | Twitter | 99 |
|  | ResearchGate | 38 |
|  | Mendeley | 3 |
| **Following** | Twitter | 130 |
|  | ResearchGate | 23 |
|  | Mendeley | 2 |

Population:
GSC (n = 811); ResearchGate (n = 515); Researcher ID (n = 275); Twitter (n = 226); Mendeley (n = 185).

As we can see, the median h-index in GSC for authors in our sample ($\tilde{x}=6$) is higher than the median h-index according to ResearcherID ($\tilde{x}=4$). This is most likely a consequence of the higher document coverage in Google Scholar (GS) as compared to the Web of Science. This is also visible if we look at the total number of citations received. The median value according to GS is $\tilde{x}=156$, almost twice the median value according to ResearchGate ($\tilde{x}=85$), which is still higher than the median value of citations according to ResearcherID ($\tilde{x}=63$). Regarding the raw total number of publications, this information was not readily available in GSC profiles, so we could only extract it from ResearchGate (median $\tilde{x}=27$), ResearcherID ($\tilde{x}=15$), and Mendeley ($\tilde{x}=9$). In terms of social interaction metrics, Twitter is the platform that contains more information about followers and followees. It is also worth noting that ResearchGate accumulates more information about social interactions in its platform than Mendeley.

It is important to know that in some cases, even when a profile has been created in one of these platforms, some of the indicators that the platform usually provides are not available. This might be caused by a number of reasons. Table 7 shows the number of profiles in which each metric was available, equal to zero, or not available. The platforms that presented a larger number of profiles in which metrics were unavailable were ResearcherID and Mendeley. ResearchGate also presented a large number of profiles in which the RG Score, the total number of citations, and the Impact Points indicator were not available. In GSC, 25% of the sample had an i-index of 0. This is because these authors did not have any document with at least 10 citations.

**Table 7.**
Number of authors for whom metrics are either or not available in each of the social platforms.

| PLATFORM | METRIC | Available | Equal to zero | Not available | % (zero) | % (av.) |
|---|---|---|---|---|---|---|
| GSC | Citations | 811 | 42 | 0 | 5.2 | 0 |
| | h-index | 811 | 42 | 0 | 5.2 | 0 |
| | i10 index | 811 | 203 | 0 | 25.0 | 0 |
| | Citations (Last 5 years) | 811 | 46 | 0 | 5.7 | 0 |
| | h-index (Last 5 years) | 811 | 46 | 0 | 5.7 | 0 |
| | i10 index (Last 5 years) | 811 | 216 | 0 | 26.6 | 0 |
| RESEARCHER ID | Total Articles | 328 | 113 | 483 | 34.5 | 59.5 |
| | Articles with Citation Data | 328 | 131 | 483 | 39.9 | 59.5 |
| | Sum of the Times Cited | 328 | 140 | 483 | 42.7 | 59.5 |
| | Average Citations per Item | 328 | 140 | 483 | 42.7 | 59.5 |
| | h-index | 328 | 140 | 483 | 42.7 | 59.5 |
| RESEARCH GATE | RG Score | 543 | 61 | 268 | 11.2 | 33..0 |
| | Publications | 543 | 28 | 268 | 5.2 | 33..0 |
| | Views | 543 | 22 | 268 | 4.1 | 33..0 |
| | Downloads | 543 | 40 | 268 | 7.4 | 33..0 |
| | Citations | 543 | 56 | 268 | 10.3 | 33..0 |
| | Impact Points | 543 | 118 | 268 | 21.7 | 33..0 |
| | Profile views | 543 | 3 | 268 | 0.6 | 33..0 |
| | Following | 543 | 29 | 268 | 5.3 | 33..0 |
| | Followers | 543 | 16 | 268 | 2.9 | 33..0 |
| MENDELEY | Readers | 334 | 156 | 477 | 32.7 | 58.8 |
| | Publications | 334 | 149 | 477 | 31.2 | 58.8 |
| | Readers per document | 185 | 7 | 626 | 3.8 | 77.2 |
| | Followers | 334 | 122 | 477 | 25.6 | 58.8 |
| | Following | 334 | 156 | 477 | 32.7 | 58.8 |
| TWITTER | Tweets | 240 | 14 | 571 | 5.8 | 70.4 |
| | Followers | 240 | 1 | 571 | 0.4 | 70.4 |
| | Following | 240 | 3 | 571 | 1.3 | 70.4 |
| | Days | 240 | 0 | 571 | 0 | 70.4 |
| | Sum Retweets | 240 | 82 | 571 | 34.2 | 70.4 |
| | H-Retweets | 240 | 82 | 571 | 34.2 | 70.4 |

Available: metric available in the platform (including zeroes)
Equal to zero: number of authors with the corresponding metric equal to "0"
Not av.: metric not available in the profile.
% (zero): percentage of profiles in which the metric is ZERO, respect to the total number of authors with a profile in the corresponding platform
% (tot): percentage of profiles in which the metric is NOT available, respect to the total number of authors in our sample (811).

### 3.3. From citations to followers: a comparison of academic profile metrics

All the metrics displayed by GSC correlate strongly with one another, which makes sense because all of them are based on citations (Figure 3). In ResearchGate, however, we find a clear separation between usage (views and downloads) and citation-based metrics (total received citations, impact points, RG Score), and social interaction indicators. This separation can be observed in several platforms. Additionally, moderate to very high correlations are found between citation-based indicators in different platforms, i.e. correlations between the RG Score and all indicators in GSC are very high, whereas metrics in ResearcherID correlate only moderately with GSC and RG metrics, maybe because metrics in ResearcherID profiles are only updated when the user updates his/her profile, and therefore the data we collected was probably outdated.

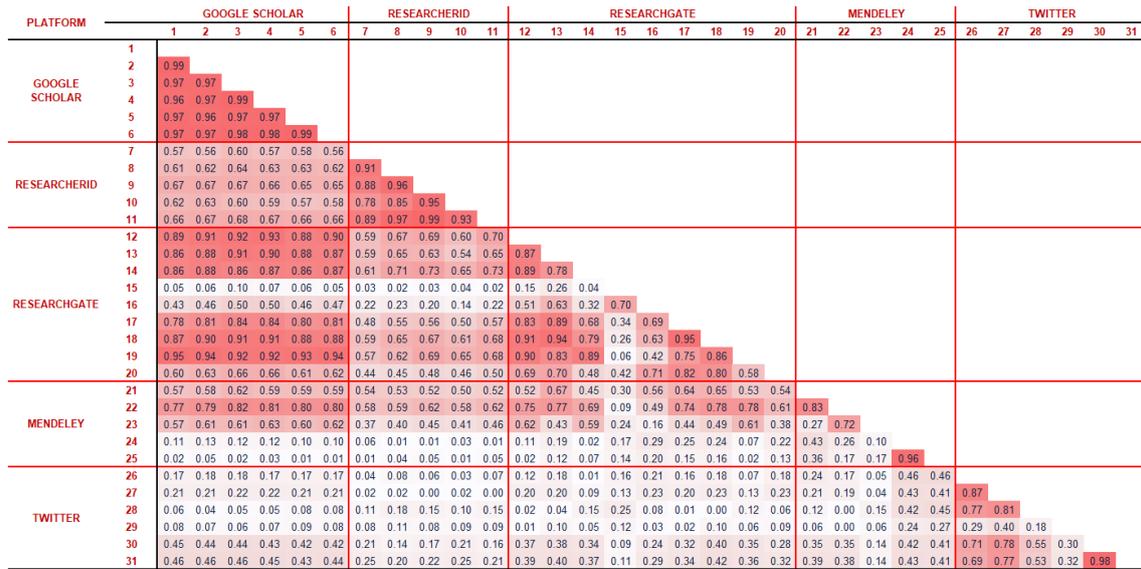

**Fig 3.** Correlation matrix (Spearman) for 31 social platform profile metrics associated with the bibliometrics community.

| | GSC | | ResearcherID | | ResearchGate | | Mendeley | | Twitter |
|---|---|---|---|---|---|---|---|---|---|
| 1 | Citations | 7 | Articles | 12 | RG Score | 17 | Downloads | 21 | Publications | 26 | Tweets |
| 2 | Citations (5Y) | 8 | Articles cited | 13 | Publications | 18 | Views | 22 | Readers | 27 | Followers |
| 3 | h-index | 9 | Sum of times cited | 14 | Impact Points | 19 | Citations | 23 | Readers/doc | 28 | Following |
| 4 | h5-index | 10 | Average citations | 15 | Following | 20 | Profile Views | 24 | Followers | 29 | Days |
| 5 | i10-index | 11 | h-index | 16 | Followers | | | 25 | Following | 30 | Sum reTweets |
| 6 | i10-index (5Y) | | | | | | | | | 31 | H reTweets |

The number of readers in Mendeley exhibits a very particular behaviour. While it correlates with the usage metrics offered by ResearchGate, it also achieves moderately high correlations with Google Scholar's total citations (r= 0.77) and h-index (r= 0.82), and with the RG Score (r= 0.75).

Regarding Twitter, Figure 3 shows that the number of tweets published, and the number of followers or followees do not correlate well with metrics from other platforms, but only among themselves. Only the sum of retweets and the H retweets have a moderate correlation with citation-based metrics from other platforms (r= 0.44 for total citations in GSC and sum of retweets, and r= 0.45 for total citations in GSC and h-retweets). However, the number of days that a Twitter account has been active (a variable included to check whether it may influence other Twitter metrics) does not seem to correlate with any other metric, not even with the other metrics extracted from Twitter. This suggests that time (whether an author is veteran or rookie in the platform) is not a critic factor to achieve a high number of followers.

Perhaps surprisingly, follower counts across different platforms do not seem to correlate well. Correlations between follower counts in ResearchGate and Twitter (0.23), and ResearchGate and Mendeley (0.29) can only be considered to be low. Only between Twitter and Mendeley did we find a correlation that could be considered moderate (0.43). What it is clear is that connectivity metrics, such as follower counts, do not correlate well at all with citation-based metrics. This separation can be visualized through the Principal Component Analysis (PCA) available in Figure 4.

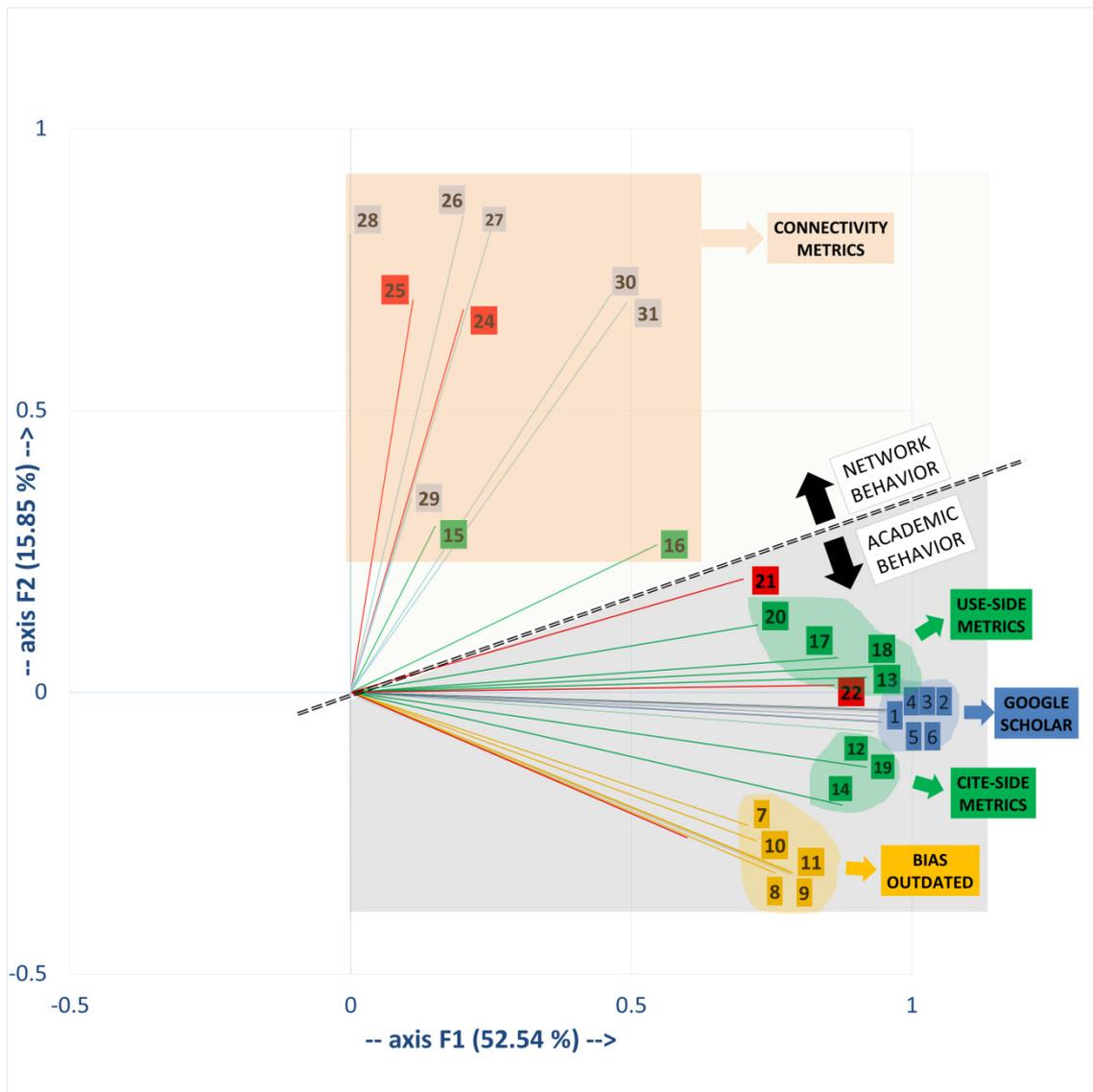

**Fig. 4.** Principal Component Analysis (PCA) for 31 author-level metrics in the bibliometrics community.

| | **GSC** | | **ResearcherID** | | **ResearchGate** | | **Mendeley** | | **Twitter** |
|---|---|---|---|---|---|---|---|---|---|
| 1 | Citations | 7 | Articles | 12 | RG Score | 17 | Downloads | 21 | Publications | 26 | Tweets |
| 2 | Citations (5Y) | 8 | Articles cited | 13 | Publications | 18 | Views | 22 | Readers | 27 | Followers |
| 3 | h-index | 9 | Sum of times cited | 14 | Impact Points | 19 | Citations | 23 | Readers/doc | 28 | Following |
| 4 | h5-index | 10 | Average citations | 15 | Following | 20 | Profile Views | 24 | Followers | 29 | Days |
| 5 | i10-index | 11 | h-index | 16 | Followers | | | 25 | Following | 30 | Sum reTweets |
| 6 | i10-index (5Y) | | | | | | | | | 31 | H reTweets |

The existence of differences in correlations according to the nature of metrics (total, average, and hybrid scores) may introduce a bias in the previous PCA (Figure 4). In this work, out of the 31 metrics included in the PCA, only 3 are average metrics and 6 hybrid metrics (the remaining 22 are total metrics), minimizing the effect. By way of illustration, a PCA composed only by the 6 hybrid metrics can be observed in Figure 5, reinforcing in this case the previous results. As we can see, while h-index values from Google Scholar are related with RG score and Impact Points in ResearchGate, the h-index from Mendeley is slightly separated. The h-index of ReTweets clearly shows a different impact dimension instead.

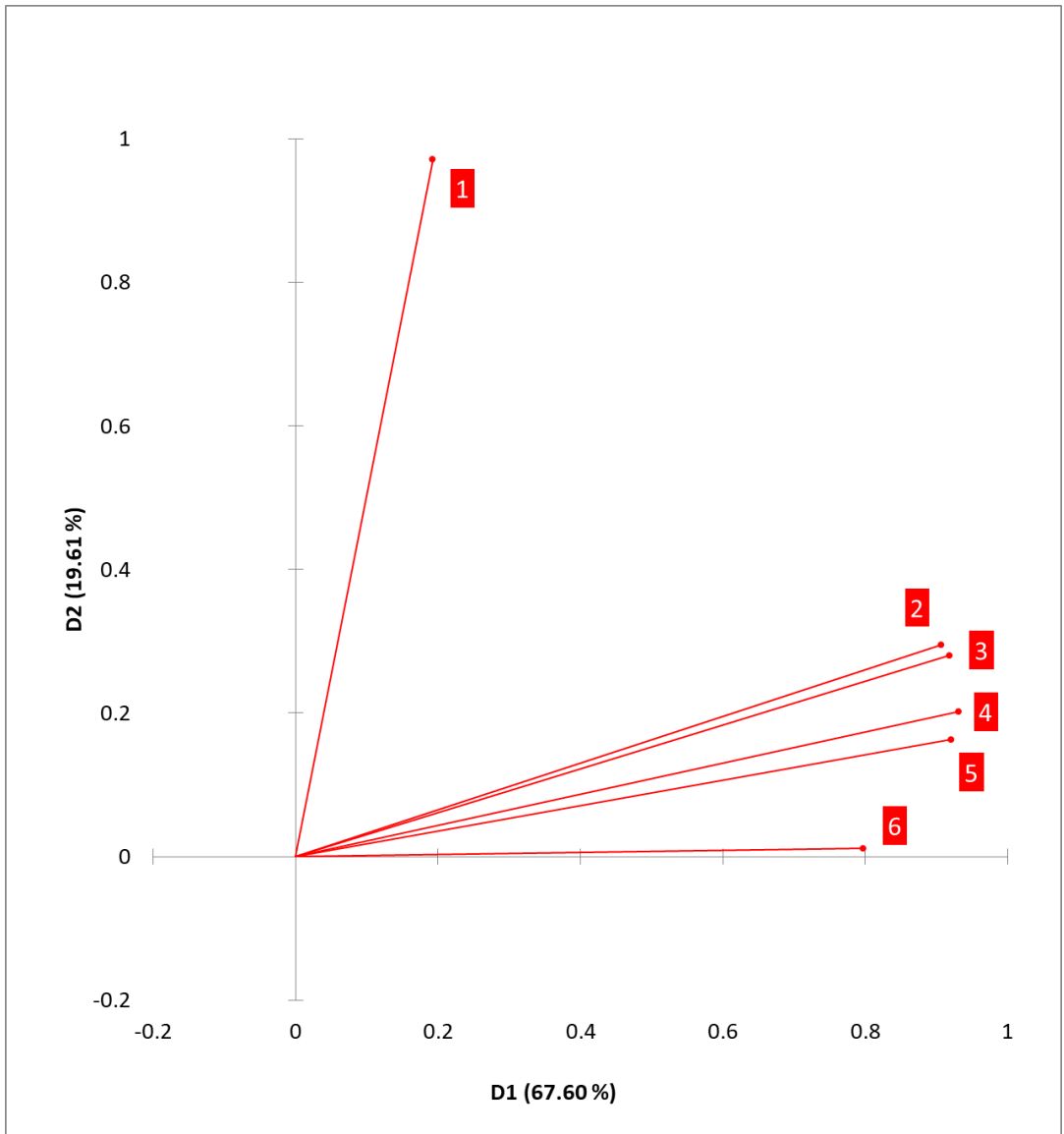

**Fig. 5.** Principal Component Analysis (PCA) for 6 hybrid author-level metrics in the bibliometrics community.
1: **Twitter**_H ReTweets; 2: **GS**_H-index; 3: **GS**_H5 index; 4: **RG**_Score; 5: **RG**_Impact Points; 6: **ResearcherID**_H-Index

Lastly, the PCA related to the 22 total metrics is offered in Figure 6. Although some logical differences with Figure 4 can be observed, the main findings remain unaltered: following/follower metrics form one separated dimension, and citation-based metrics form another one. Mendeley (Readers) is located halfway between the dimentions mentioned above. In this case, however, the number of citations from ResearcherID seems to be located near citation-like metrics, confirming on the one hand that citation metrics make up a dimensions of their own, and on the other hand, the dependence on platforms' coverage and updating mechanisms.

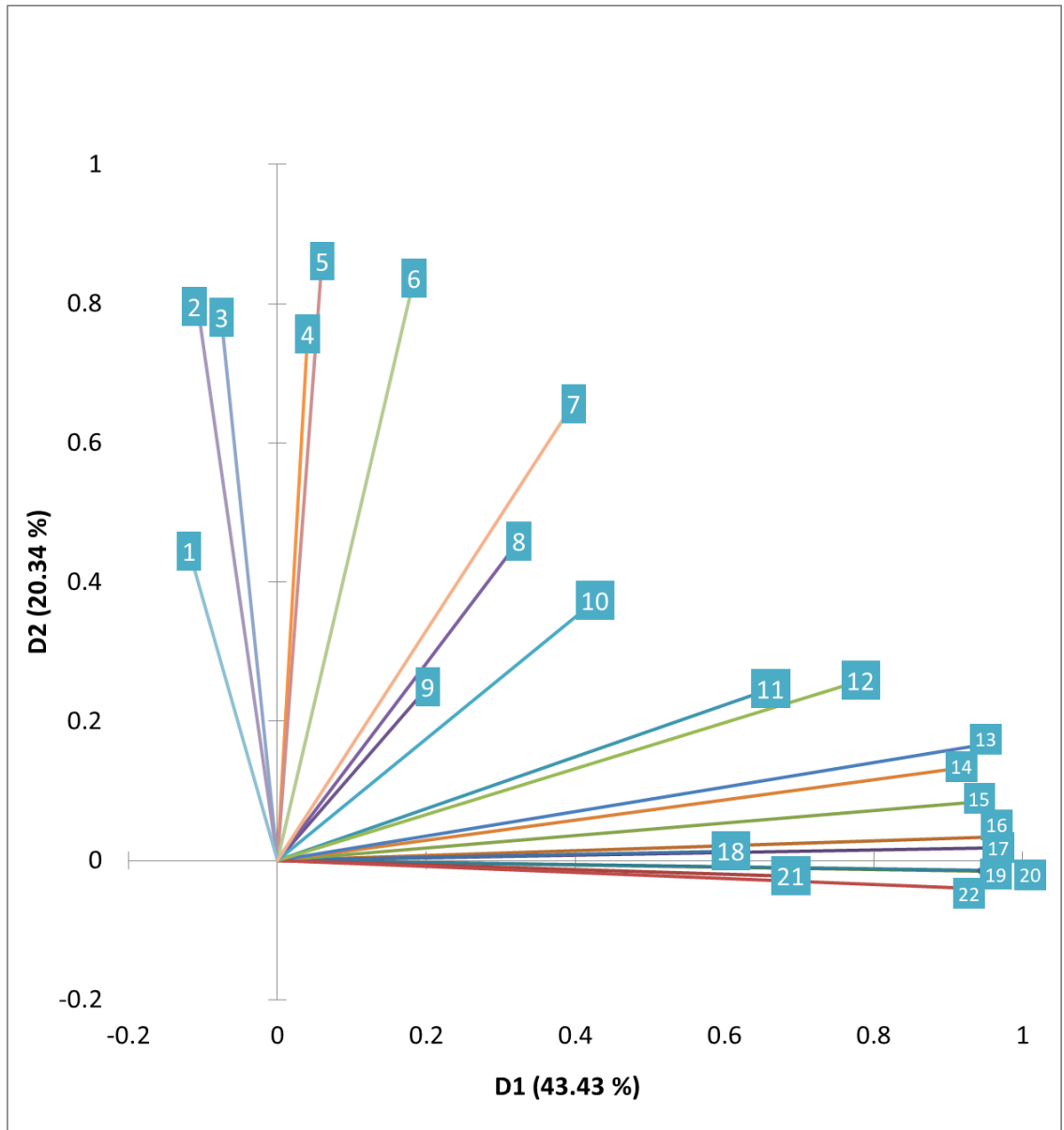

**Fig. 6.** Principal Component Analysis (PCA) for 22 total author-level metrics in the bibliometrics community.
1: **Twitter**_Days; 2: **Twitter**_Following; 3: **Mendeley**_Following; 4: **Mendeley**_Followers; 5: **Twitter**_Tweets; 6: **Twitter**_Followers; 7: **Twitter**_Sum ReTweets; 8: **Mendeley**_Publications; 9: **RG**_Following; 10: **Mendeley**_Readers; 11: **RG**_Followers; 12: **RG**_Profile Views; 13: **RG**_Views; 14: **RG**_Downloads; 15: **RG**_Publications; 16: **GS**_i10-index(5Y); 17: **GS**_Citations (5Y); 18: **ResearcherID**_Total articles; 19: **GS**_i10-index; 20: **ResearcherID**_Sum of times cited 21: **GS**_Citations; 22: **RG**_Citations;

## 4. Discussion

One of the main limitations of the sample of authors used in this study is the fact that it does not consider all the researchers in the field, because selection was dependent on having created and made public a profile in Google Scholar Citations. Moreover, although there are evidences that suggest that Google Scholar Citations was the most popular profile service in the field under study (Haustein et al., 2014), there could be some researchers in the area who have not created a profile in GSC, but have profiles in other platforms. Our sample would also miss these cases. For this reason, we acknowledge that the sample has a strong bias towards GSC. Our methodology to find author profiles relied on manual searching, and therefore we might have missed some profiles, especially if the authors did not use the same names they use in their research. This could be especially problematic in Twitter. Future studies could make use of more elaborate methodologies for author profile detection, like the one described in Costas, van Honk, and Franssen (2017). Moreover, the number of people who use academic profiles, regardless of the platform they choose, is still only a fraction of the total number of people working in any given discipline. In spite of this, this sample (811 researchers) is the largest to date in studies that aim to analyse the discipline of bibliometrics.

Overall, the results obtained in this work regarding the level of online presence of scientists in social networks are similar to the results found by Van Noorden (2014), and Bosman and Kramer (2016). However, the difference in the uptake of GSC and ResearchGate is higher in this work (only 67% of the authors with a GSC profile also have a ResearchGate profile). The most likely reason for this is that this work only studies bibliometrics researchers, which might have a stronger preference for GSC in detriment of ResearchGate. Previously, Haustein et al. (2014) had found more similar figures in the uptake of these two platforms among bibliometrics researchers. Also, the data collected by Kramer and Bosman (2016) show that ResearchGate and Google Scholar Citations are the most used profile platforms by the respondents to their survey, with a small advantage of ResearchGate over GSC.

Time is also an important factor to consider, because as Haustein et al. (2014) already found, there was a significant increase in the number of researchers who had a profile in the relatively short period of time between the collection of their two samples of data. In the past two years, ResearchGate has emerged as one of the most popular options for researcher to create an academic profile. It is therefore possible that the percentage of researchers that currently have a profile in these platforms has changed since the data for this analysis was collected. There are already studies that show the increasing preference of ResearchGate over Google Scholar Citations, like the already mentioned survey by Kramer and Bosman (2016), and also Mikki et al. (2015), who found ResearchGate to be the platform with more profiles of researchers from the University of Bergen (76% had a profile in this platform).

Regarding the use of multiple profile by the same researchers, the results in this study differ from those found in Mikki et al. (2015) and Ortega (2015a). While Mikki et al. found that 77% of the researchers in their sample only maintained a profile in one platform, (72% in the case of Ortega's study), our analysis found that only 19% of the researchers analysed had only one profile (in GSC). According to our results, 26% had two profiles, 23% three, 20% four, and 11% had a profile in the five platforms that we analysed. The differences might be explained by the particular circumstances of our sample of researchers, who may be more naturally inclined to explore these tools because they are closely related to their field of study. These differences notwithstanding, the results in this study match the finding in Mikki et al. (2015) and Ortega (2015a) that the most common pairwise combination of profiles for authors to have is the combination of GSC and ResearchGate.

*RQ 1: Data volumes*

It is important to note that some of the metrics analysed in this study are no longer available from the platforms that previously displayed them. This is the case of the Views and Downloads metrics in ResearchGate, which were combined into the Reads[5] metric shortly after we collected our data. The Impact Points metric was also hidden, in this case without giving a public explanation. Conversely, some platforms have added new metrics to their portfolio. ResearchGate started computing the h-index (with and without self-citations). Mendeley also updated its Stats page, which for a long time was also visible to the owner of the profile, giving the choice to make it public to everyone. This stats page displays Media mentions (powered by Newsflo), the h-index and total number of citations received (powered by Scopus), the total number of times an author's papers have been viewed, according to ScienceDirect, and of course, the total number of Mendeley readers.

There are several limitations that complicate making comparisons between the author-level and article-level indicators calculated by each platform (e.g. number of people with a profile, number of citing documents found by the platform, and the actual level of interaction among scientists in the platform). Nevertheless, there are significant differences in similar indicators depending on the source of the data: the median number of citations received by authors according to GSC is twice that of RG; the median number of total publications published by authors according to RG is twice that of Mendeley (see Table 6). Lastly, the median number of people that follow an author (followers) and the number of people the author follows (followees) in Twitter are higher by far than the same indicators in ResearchGate and Mendeley. Despite the methodological limitations, these results suggest that some platforms are able to provide more information than others depending on what aspect of an author's academic activities one is interested in.

*RQ 2: Author-level metrics*

The usefulness of the metrics that were obtained (which were subsequently used to compute correlations and other comparative measurements) depends on the accuracy of the data provided by each platform. GSC is known to suffer from bibliographic inconsistencies (Jacsó, 2012). It has also been reported that it introduces biases against some disciplines and institutions (Ortega, 2015b), and its citation metrics can be easily gamed (Delgado López-Cózar, Robinson-Garcia & Torres-Salinas, 2014). Regarding ResearchGate, some studies have discussed that its flagship indicator, the RG Score, is not an accurate measure of an author's authority in a scientific community, but a measure of how much the author engages in the platform itself, which also opens the metric to manipulation (Kraker and Lex, 2015; Jordan, 2015; Orduna-Malea, Martin-Martin and Delgado López-Cózar, 2016b; Orduna-Malea et al, 2017). Nevertheless, for the purposes of this specific analysis, the consequences of these issues are considered to be low, and should not have affected the correlations.

The correlations found between the different metrics reinforce previous findings obtained at the article level. Priem, Piwowar, and Hemminger (2012) also found that citation-based indicators cluster closely together. Schlögl et al. (2014) argued that downloads, Mendeley readership, and citations reflect different aspects of impact, and Glänzel and Gorraiz (2015) claimed that there are also differences between usage metrics and alternative metrics. The PCA analysis carried out in this study (Figure 4) indeed reflects these differences: Views, Downloads, and Mendeley readers are grouped together, but apart from the social interaction metrics. Therefore, these results agree with the findings by

---

[5] https://www.researchgate.net/blog/post/introducing-reads

Naude (2016), who suggested that download counts especially, but also Mendeley readership, can be a useful complement to the citation data available in GSC.

Our results also agree with previous findings (Mikki et al., 2015) when we look at author-level metrics. The indicators provided by the same platform tend to correlate, even when they reflect different aspects of impact (e.g. a strong correlation was found between downloads and citations in ResearchGate: 0.75). Additionally, citation-based metrics tend to correlate across platforms: e.g. between GSC and ResearchGate, and to a lower but still considerable degree, between the previous two and ResearcherID. The reason for the lower correlations when ResearcherID is involved has to do with the lack of information in the profiles in this platform: 34.5% of the profiles are empty, and citation indicators in this platform can only be updated manually by the owner of the profile.

The data we extracted from Twitter yielded similar results to those found by De Winter (2015), who concludes that the scientific citation process acts relatively independently of the social dynamics on Twitter. On the other hand, Ortega (2016) concludes that the number of followers indirectly influences the citation impact because participation on Twitter affects the dissemination of research papers, and therefore it may indirectly favour the likelihood of academic outputs being cited. Our findings seem to contradict Ortega's claims (no correlation between number of followers and number of citations received) but we do not think the results can disprove the claim, because correlation does not imply causation. The number and exact composition of followers is a factor that may decisively influence the degree of dissemination of academic outputs (whether the follower base is actually the target audience of the publications). This issue was not addressed in this study.

Lastly, the obtained correlations and PCA results should be taken cautiously since metrics have not been normalized. For example, the age of authors has not been controlled. In this sense, more experienced authors may exhibit a distinctive behaviour compared to emerging authors. This fact does not jeopardize the main findings of this work (identifying different web impact dimensions through raw author-level metrics provided by social networks at discipline level). However, the identification of author clusters (sharing common attributes such as similar academic age range, online activity, gender, language, etc.) constitutes a future line of research.

## 5. Conclusions

The results indicate that an important number of the researchers in our sample only had a profile in GSC (159), although many of them (543, 67%) also had a profile in ResearchGate, which made it the second most used platform at the time of data collection. The usage indicators (currently, Reads) and the networking capabilities provided by ResearchGate are features that GSC lack.

The analysis finds two main dimensions of online impact (RQ1). There is a cluster of metrics related to academic performance, which can be further subdivided into two subclusters: usage metrics (views, downloads), and citation metrics (citation counts, h-index). The other cluster contains metrics related to social connectivity and popularity (followers).

The authors in our sample seem to prefer Twitter over Mendeley and ResearchGate when it comes to engaging in social interactions. In Mendeley, the researchers in our sample attract a significant amount of followers, but do not tend to use this platform to keep informed about new publications: they mainly use Twitter for this purpose. ResearchGate seemed to be emerging as a source for researchers to keep themselves informed about the latest published research in their fields.

The data suggest that GSC is still the source that is able to provide the highest volumes of citation-related data (RQ2). Regarding social connectivity, Twitter seemed to be the platform of choice, even if it is not a purely academic platform. Nevertheless, ResearchGate also showed a significant amount of social interaction among researchers (followers/followees), much higher than in Mendeley. For this type of comparisons, it is interesting to put side-by-side similar metrics provided by different platforms. Doing this can provide insight into the different levels of uptake of the different platforms by a specific scientific community.

One general conclusion that can be extracted from this study is that despite the general preference towards some platforms (GSC and ResearchGate) there is not any platform in which all researchers in a discipline are present, or one that collects and provides the best data across all dimensions (citations, usage, social connectivity). Nevertheless, we found that publication and citation metrics correlate more or less consistently across platforms.

Lastly, these results should always be interpreted within the context of the bibliometrics community. The study of other communities might very well yield different results. For example, penetration rates of these platforms in other communities might be different, because for the bibliometrics community, academic profile platforms are an increasingly interesting object of study.

## Acknowledgments

Alberto Martín-Martín enjoys a four-year doctoral fellowship (FPU2013/05863) granted by the Ministerio de Educación, Cultura, y Deportes (Spain). We also thank Juan Manuel Ayllón for his valuable help cleaning data in the early stages of this study.